\documentclass[11pt,a4paper]{article}

\newif\ifjhepstyle
\jhepstylefalse
\jhepstyletrue

\ifjhepstyle
	\usepackage{jheppub}
	\usepackage{amsfonts}
	\usepackage{verbatim}
	\usepackage{float}
	\restylefloat{table}
	
\else	\usepackage{verbatim}
	\usepackage{cite}
	\usepackage{setspace}
	\usepackage[top=2.45cm, bottom=2.5cm, left=2.45cm, right=2.45cm]{geometry}
	\usepackage{amsmath,amsfonts,amssymb}
	\usepackage[colorlinks=true
	,urlcolor=blue
	,anchorcolor=blue
	,citecolor=blue
	,filecolor=blue
	,linkcolor=blue
	,menucolor=blue
	]{hyperref}
	\usepackage{float}
	\restylefloat{table}
	
	\numberwithin{equation}{section}
\fi
\newcommand{\ThisIsTheAbstract}{We show that recently proposed free boundary conditions for AdS$_3$ are dual to two-dimensional quantum gravity in certain fixed gauges. In particular, we note that an appropriate identification of the generator of Virasoro transformations leads to a vanishing total central charge in agreement with the theory at the boundary. We argue that this identification is necessary to match the bulk and boundary generators of Virasoro transformations and for consistency with the constraint equations.}
\ifjhepstyle
\title{Free boundary conditions and the AdS$_3$/CFT$_2$ correspondence}

\author{Luis Apolo}
\author{and Massimo Porrati}

\affiliation{Center for Cosmology and Particle Physics, Department of Physics, New York University, \\4 Washington Place, New York, NY 10003, USA}

\emailAdd{lav271@nyu.edu}
\emailAdd{massimo.porrati@nyu.edu}

\abstract{\ThisIsTheAbstract} 

\fi
\begin{document}

\ifjhepstyle
\maketitle
\flushbottom
\fi

%%%%%%%%%%%%%%%%%%%%%%%%%%%%%%%%%%%%%%%%%%%%%%%%%%%
% Macros
%%%%%%%%%%%%%%%%%%%%%%%%%%%%%%%%%%%%%%%%%%%%%%%%%%%
\long\def\symfootnote[#1]#2{\begingroup%
\def\thefootnote{\fnsymbol{footnote}}\footnote[#1]{#2}\endgroup} 

% Definitions
\def\({\left (}
\def\){\right )}
\def\lb{\left [}
\def\rb{\right ]}
\def\lB{\left \{}
\def\rB{\right \}}

\def\Int#1#2{\int \textrm{d}^{#1} x \sqrt{|#2|}}
\def\Bra#1{\left\langle#1\right|} 
\def\Ket#1{\left|#1\right\rangle}
\def\BraKet#1#2{\left\langle#1|#2\right\rangle} 
\def\Vev#1{\left\langle#1\right\rangle}
\def\Vevm#1{\left\langle \Phi |#1| \Phi \right\rangle}\def\bbox{\bar{\Box}}
\def\til#1{\tilde{#1}}
\def\wtil#1{\widetilde{#1}}
\def\ph#1{\phantom{#1}}

\def\ra{\rightarrow}
\def\la{\leftarrow}
\def\lra{\leftrightarrow}
\def\p{\partial}
\def\diff{\mathrm{d}}

\def\sinh{\mathrm{sinh}}
\def\cosh{\mathrm{cosh}}
\def\tanh{\mathrm{tanh}}
\def\coth{\mathrm{coth}}
\def\sech{\mathrm{sech}}
\def\csch{\mathrm{csch}}

\def\a{\alpha}
\def\b{\beta}
\def\g{\gamma}
\def\d{\delta}
\def\e{\epsilon}
\def\ve{\varepsilon}
\def\k{\kappa}
\def\l{\lambda}
\def\n{\nabla}
\def\s{\sigma}
\def\t{\theta}
\def\z{\zeta}
\def\vp{\varphi}

\def\ss{\Sigma}
\def\dd{\Delta}
\def\gg{\Gamma}
\def\ll{\Lambda}
\def\tt{\Theta}

\def\D{{\cal D}}
\def\F{{\cal F}}
\def\H{{\cal H}}
\def\J{{\cal J}}
\def\K{{\cal K}}
\def\L{{\cal L}}
\def\O{{\cal O}}
\def\P{{\cal P}}
\def\W{{\cal W}}
\def\X{{\cal X}}
\def\Z{{\cal Z}}

\def\zz{\bar z}
\def\xx{\bar x}
\def\xp{x^{+}}
\def\xm{x^{-}}

\def\VirU1{\mathrm{Vir}\otimes\hat{\mathrm{U}}(1)}
\def\VirSL2R{\mathrm{Vir}\otimes\widehat{\mathrm{SL}}(2,\mathbb{R})}
\def\U1{\hat{\mathrm{U}}(1)}
\def\SL2R{\widehat{\mathrm{SL}}(2,\mathbb{R})}
\def\sl2r{\mathrm{SL}(2,\mathbb{R})}
\def\by{\mathrm{BY}}
%%%%%%%%%%%%%%%%%%%%%%%%%%%%%%%%%%%%%%%%%%%%%%%%%%%

%%%%%%%%%%%%%%%%%%%%%%%%%%%%%%%%%%%%%%%%%%%%%%%%%%%
% Plain style titlepage and ToC
%%%%%%%%%%%%%%%%%%%%%%%%%%%%%%%%%%%%%%%%%%%%%%%%%%%
\unless\ifjhepstyle
\begin{titlepage}
\begin{center}

\ph{.}

\vskip 4 cm

{\Large \bf Free boundary conditions and the AdS$_3$/CFT$_2$ correspondence}

\vskip 1 cm

{Luis Apolo and Massimo Porrati}

\vskip .75 cm

{\em Center for Cosmology and Particle Physics, \\Department of Physics, New York University, \\4 Washington Place, New York, NY 10003, USA}

\end{center}

\vskip 1.25 cm
%\title{}
%\author{}
%\date{}Killing
%\maketitle

\begin{abstract}
\noindent \ThisIsTheAbstract
\end{abstract}
\end{titlepage}
\newpage
\fi

\unless\ifjhepstyle
\tableofcontents
\noindent\hrulefill
\bigskip
\fi
%%%%%%%%%%%%%%%%%%%%%%%%%%%%%%%%%%%%%%%%%%%%%%%%%%%

%%%%%%%%%%%%%%%%%%%%%%%%%%%%%%%%%%%%%%%%%%%%%%%%%%%

%%%%%%%%%%%%%%%%%%%%%%%%%%%%%%%%%%%%%%%%%%%%%%%%%%%
\section{Introduction and summary}
%%%%%%%%%%%%%%%%%%%%%%%%%%%%%%%%%%%%%%%%%%%%%%%%%%%

A fundamental aspect of the AdS/CFT correspondence~\cite{Maldacena:1997re,Witten:1998qj,Gubser:1998bc} is that the isometries of spacetime in the bulk must match global symmetries of the theory at the boundary. Indeed, the symmetry group of AdS$_{d+1}$, SO$(d,2)$, is precisely that of a conformal field theory in $d$ dimensions when $d > 2$. The case $d = 2$ is special in that the SO$(2,2)$ symmetry of the dual theory is enhanced to an infinite number of local symmetries described by two copies of the Virasoro algebra. As shown by Brown and Henneaux~\cite{Brown:1986nw}, these symmetries are realized asymptotically in AdS$_3$ by imposing Dirichlet boundary conditions where the boundary metric is non-dynamical, i.e.
  \begin{align}
  \mathrm{ds}^2_{r\ra\infty} = r^2 \mathrm{ds}^2_{b}, && \mathrm{ds}^2_{b} = -\diff x^+ \diff x^-,
  \end{align}
where $\mathrm{ds}^2_{b}$ is the line element at the boundary and $x^{\pm} = t \pm \phi$ are lightcone coordinates.

The boundary conditions found by Brown and Henneaux are the most general set of Dirichlet boundary conditions that lead to finite charges in pure three-dimensional gravity with a negative cosmological constant. It is possible to obtain new boundary conditions by promoting the boundary metric to a dynamical variable and making its conjugate momentum, the Brown-York stress-energy tensor, vanish~\cite{Compere:2008us}. This amounts to imposing Neumann boundary conditions in the 3D theory,
  \begin{align}
  \mathrm{ds}^2_{b} = - \g_{ij}\diff x^i \diff x^j, && T_{ij} = \frac{4\pi}{|\g|^{1/2}} \frac{\d S_G}{\d \g^{ij}} = 0. \label{eq:neumannbc}
  \end{align}
Here $S_G$ is the total gravitational action\footnote{$S_G$ contains the Gibbons-Hawking term and appropriate boundary counterterms necessary for finiteness of the action and vanishing of the Brown-York tensor.}, $\g_{ij}$ is the metric at the boundary, and $T_{ij}$ is the Brown-York stress-energy tensor~\cite{Brown:1992br,Balasubramanian:1999re}. Making the boundary metric dynamical means that we must now integrate over $\g_{ij}$ in the partition function of the dual theory. Thus, the theory at the boundary is also a theory of gravity and the vanishing of the Brown-York tensor in the bulk is a consequence of the equations of motion of $\g_{ij}$. In particular, the diffeomorphism invariance of the dual theory allows us to choose a gauge for the boundary metric. Then the non-vanishing components of the Brown-York stress-energy tensor are to be interpreted as constraints on the physical states of the dual theory.

Recently, new boundary conditions for AdS$_3$ were introduced where the boundary metric is in a {\it chiral lightcone} gauge~\cite{Compere:2013bya}, 
  \begin{align}
  \mathrm{ds}^2_{b} = - \diff x^+ \lb \diff x^- + \rho(x^+)\diff x^+ \rb, 
  \end{align}
a conformal gauge~\cite{Troessaert:2013fma},
  \begin{align}
  \mathrm{ds}^2_{b} = - e^{2\vp(x)} \diff x^+ \diff x^-, \label{eq:conformalgauge}
  \end{align}
or a lightcone gauge~\cite{Avery:2013dja},
  \begin{align}
  \mathrm{ds}^2_{b} = - \diff x^+ \lb \diff x^- + \mu(x) \diff x^+ \rb. \label{eq:lightconegauge}
  \end{align}
Along with these new boundary conditions come new asymptotic symmetries for AdS$_3$ (see Table~\ref{theonlytable}) where the Virasoro central charge is $c = 3l/2G$, $l$ is the AdS radius and $G$ is Newton's constant, while the Kac-Moody level is $k = -c/6$\footnote{However note that for the $\U1$ symmetries it is always possible to change the value of the level by normalization of the currents but not its sign. In particular for the boundary conditions of ref.~\cite{Compere:2013bya} the normalization depends on the background.}.
  \begin{table}[H]
  \begin{center}
  \begin{tabular}{lclcl}
  Gauge & & Symmetry & \\ \hline 
  Chiral lightcone & \hspace{2mm} & $\VirU1$ & \\
  Conformal & \hspace{2mm} & $( \VirU1 )^2$ & \\
  Lightcone & \hspace{2mm} & $\VirSL2R$ & 
  \label{theonlytable}
  \end{tabular}
  \caption{Symmetries of new AdS$_3$ boundary conditions.}
  \label{symmetries}
  \end{center}
  \end{table}

In this paper we study in more detail the boundary conditions of refs.~\cite{Troessaert:2013fma} and~\cite{Avery:2013dja}. We show that AdS$_3$ with these boundary conditions is dual to 2D quantum gravity in either the conformal or lightcone gauges. The dual theory is formally described by the partition function
  \begin{align}
  \Z = \int \D\phi_a \D\hat\g\, \mathrm{exp}\,i\lB  S_{m+g}[\phi_a,\hat\g] + \frac{k}{16\pi} S_{p}[\hat\g]\rB, \label{eq:dualtheory}
  \end{align}
where $\phi_a$ represents the matter \emph{and} ghost fields that result from gauge fixing and $S_{m+g}$ is the sum of their actions. The central charge of the matter and ghost systems is denoted by $c_{M} \ne 0$ which implies that the matter plus ghost theory, assumed to be conformally-invariant at the classical level, has a non-vanishing Weyl anomaly. Here $\hat\g$ represents the gravitational degree of freedom that is not fixed by the diffeomorphism invariance of the theory, i.e. $\hat \g = \vp$ in the conformal gauge and $\hat \g = \mu$ in the lightcone gauge\footnote{Note that in the conformal gauge the action $S_{m+g}[\phi_a]$ does not depend on the conformal factor $\vp$ since it couples to the trace of the stress-energy tensor which we have assumed to vanish at the classical level.}. 

The term denoted by $S_p$ is the Polyakov action~\cite{Polyakov:1987zb}
  \begin{align}
  S_p = \int \sqrt{|\g|} \diff^2x \( R \frac{1}{\Box} R + \l \),
  \end{align}
where $\l$ is a cosmological constant which will be set to zero in the remainder of this paper\footnote{Therefore, we will confine our analysis to boundaries with the topology of
either the cylinder or the torus.}. In the conformal gauge this is possible only when the reference metric is flat, as in eq.~\eqref{eq:conformalgauge}, since the Polyakov action reduces to the Liouville action where $\vp$ is conformally-coupled. The Polyakov action is required for consistency of the theory~\cite{Polyakov:1988ok, David:1988hj, Distler:1988jt, D'Hoker:1990md}. It guarantees that the Ward identities of the theory, which are sensitive to anomalous contributions from the measure, are realized at the classical level. In particular, it guarantees that the stress-energy tensor $T_{\mu\nu}$ is covariantly conserved and that its trace reproduces the Weyl anomaly. Then
  \begin{align}
  0 = T_{\mu\nu} \equiv \frac{4\pi}{\sqrt{\g}} \frac{\d S}{\d \g^{\mu\nu}}, \label{eq:boundarytmunu}
  \end{align}
where $S = S_{m+g} + (k/16\pi) S_p$, yields the equations of motion for $\g_{\mu\nu}$, that 
are also the constraints on the physical states of the theory. Consistency of eq.~\eqref{eq:boundarytmunu} at the quantum level, where the measures of $\phi_a$ and $\hat \g$ contribute to the expectation value of $T_{\mu}^{\mu}$, requires that the total central charge of the matter, ghost, and Polyakov actions vanishes. In the semiclassical limit where $c_M$ is large the central charge of the Polyakov action is $c_p = 6k$ where $k = -c_M/6$ is the parameter in front of the Polyakov action. At the quantum level both of these quantities receive corrections of $\O(1)$ such that the total central charge $c_{total} = c_M + c_p$ always vanishes.

Thus, a consistent set of free boundary conditions for AdS$_3$ requires a vanishing central charge for the Virasoro algebra. We will show that a proper identification of the generator of Virasoro transformations yields the desired result\footnote{In the conformal gauge this result was first obtained in ref.~\cite{Troessaert:2013fma} but its interpretation was not considered there. Instead, the Brown-Henneaux central charge was recovered by shifting the generators of Virasoro transformations. We now understand this as a consequence of \emph{untwisting} the twisted Sugawara tensor.}. Thus, if we denote by $Q_{\e}$ the charge corresponding to the asymptotic Killing vector $\e$, the commutator of Virasoro charges reads 
  \begin{align}
  i[Q_{\e},Q_{\s}] = Q_{\e'\s-\e\s'}. \label{eq:virasorocommutator}
  \end{align}
We will show that the charges generating the Virasoro and Kac-Moody transformations match those of the dual theory at the boundary. In particular, the charge responsible for (left-moving) Virasoro transformations is given by
  \begin{align}
  Q[\e] = -\frac{1}{2\pi} \int \diff\phi \e(x^+) \lb L(x^+) + T_s(x^+) \rb, \label{eq:virasorocharge0}
  \end{align}
and a similar expression exists for the right-moving charge found in the boundary conditions of ref.~\cite{Troessaert:2013fma}. In eq.~\eqref{eq:virasorocharge0} we identify $L$ with the expectation value of the left-moving generator of Virasoro transformations on a state of the matter and ghost systems, while $T_s$ corresponds to the \emph{twisted} Sugawara tensor of the Kac-Moody algebra. In the semiclassical limit probed by semiclassical gravity in the bulk the twisted Sugawara tensor is given by
  \begin{align}
  T_s = \frac{1}{2k} \eta^{ab} J_a J_b + \p_+ J_0,  \label{eq:twistedsugawara}
  \end{align}
where $J_a$ are the currents and $\eta_{ab}$ is the Cartan-Killing metric of the corresponding current algebra. In the lightcone gauge this result is consistent with the constraints imposed on the $T_{++}$ and $T_{+-}$ components of the stress-energy tensor, since the latter reads
  \begin{align}
  0 = T_{+-} \propto J_{-}.
  \end{align}
This implies that $J_-$ is an operator of weight zero under Virasoro transformations. 

Note that it is always possible to shift $L$ by a left-moving function of the gravitational degrees of freedom. Thus, the appearance of a twisted Sugawara tensor in eq.~\eqref{eq:twistedsugawara}, as opposed to the standard Sugawara tensor, may seem artificial even though it is consistent with the transformation of the Kac-Moody currents and the current algebra. The ambiguity is fixed by matching the Brown-York tensor with the stress-energy tensor of the dual theory at the boundary. This determines the subleading components of the metric and fixes the form of the Virasoro charges. Alternatively, we can proceed without making any reference to the dual theory by computing the Brown-York stress-energy tensor \emph{off-shell} in a sense that shall be made precise in Section~\ref{se:conformalgauge}.

It is interesting to note that both $L$ and $T_s$ generate two independent Virasoro algebras, which suggests that $L$ and the Kac-Moody currents $J_a$ commute in the dual theory. We show that it is possible to recover this result in the bulk by requiring that the transformation of the subleading components of the metric is consistent with the transformation of the leading components $\vp$ and $\mu$. We then find that, under Kac-Moody transformations parametrized by $\ll(x)$,
  \begin{align}
  \d_{\ll} L = 0, 
  \end{align}
which implies that $[L, J_a] = [L, T_s] = 0$. On the other hand under Virasoro transformations parametrized by $\e(x^+)$ we have 
  \begin{align}
  \d_{\e} L = 2 \e' L + \e L' + \frac{c}{12} \e''',
  \end{align}
where $c = 3l/2G$. Thus, we find that in the semiclassical limit the central charge of the matter plus ghost systems is given by the Brown-Henneaux central charge,
  \begin{align}
  c_M = 3l/2G,
  \end{align}
and, using eq.~\eqref{eq:virasorocommutator}, that the central charge of the Polyakov action is given by
  \begin{align}
  c_p = - c_M,
  \end{align}
which is consistent with the dual theory at the boundary.

Further evidence for the correspondence between AdS$_3$ with free boundary conditions and 2D quantum gravity is obtained by studying the asymptotic behavior of a massive scalar field in these backgrounds. We find that under Virasoro transformations the weights of scalar operators vanish, in contrast with the case of AdS$_3$ with Brown-Henneaux boundary conditions. This result is consistent with the fact that the constraint equations~\eqref{eq:boundarytmunu} restrict all physical states to be singlets under Virasoro transformations. Likewise, it is not difficult to show that the Kac-Moody weights $\l$ of scalar operators are equal to their conformal weights $h$ in the absence of gravity. This result agrees with that found in refs.~\cite{Polyakov:1987zb,David:1988hj, Distler:1988jt} in the semiclassical limit where $k$ is large,
  \begin{align}
  \l = h + \frac{\l(1-\l)}{k+2}.
  \end{align}

The remainder of this paper is devoted to justifying these statements. We begin in Section~\ref{se:conformalgauge} with the boundary conditions of ref.~\cite{Troessaert:2013fma} where the boundary metric is in the conformal gauge. There we show that the Brown-York stress-energy tensor, currents, and central charges match the corresponding quantities of the theory at the boundary.  The lessons learned in the conformal gauge are then applied to the lightcone gauge. In Section~\ref{se:lightconegauge} we introduce a slight modification of the boundary conditions given in ref.~\cite{Avery:2013dja} and find the corresponding Brown-York tensor, currents, and central charges. In Section~\ref{se:lightconecft} we study general properties of conformal field theories in the lightcone gauge and show that the quantities computed in the bulk match those of the theory at the boundary. We present our conclusions in Section~\ref{se:conclusions}.

%%%%%%%%%%%%%%%%%%%%%%%%%%%%%%%%%%%%%%%%%%%%%%%%%%%
\section{Conformal gauge} 
%%%%%%%%%%%%%%%%%%%%%%%%%%%%%%%%%%%%%%%%%%%%%%%%%%%
\label{se:conformalgauge}

\subsection{Boundary conditions and conserved charges}

Let us begin with the boundary conditions of ref.~\cite{Troessaert:2013fma} where the boundary metric is in the conformal gauge and the asymptotic symmetry group consists of two independent copies of $\VirU1$. The boundary conditions are given by
  \begin{align}
  & g_{rr} = r^{-2} + \O(r^{-4}),  & g_{+-} &= -e^{2\vp(x)} r^2/2 + \O(r^0),    \notag      \\
  & g_{r\pm} = \O(r^{-3}),           & g_{\pm \pm} &= \O(r^0),   \label{eq:bc1}
  \end{align}
and correspond to a generalization of Brown-Henneaux boundary conditions~\cite{Brown:1986nw}. As discussed in ref.~\cite{Troessaert:2013fma}, a well-defined variational principle where $\vp$ is allowed to vary requires
  \begin{align}
  \p_+ \p_- \vp = 0,  \label{eq:harmoniceq}
  \end{align}
which is equivalent to the vanishing of the trace of the Brown-York stress-energy tensor. In contrast with ref.~\cite{Troessaert:2013fma} we do not impose eq.~\eqref{eq:harmoniceq} as a boundary condition but as an equation of motion. Furthermore, since we perform our analysis \emph{off-shell}, i.e. with arbitrary $\vp$, we can unambiguously determine the subleading components of the metric.

The most general solution obeying these boundary conditions may be written using Fefferman-Graham coordinates,
  \begin{align}
  \diff s^2 = g_{ab} \diff x^{a}\diff x^{b} = \frac{\diff r^2}{r^2} + \( r^2 g^{(0)}_{\mu\nu} + g^{(2)}_{\mu\nu} + r^{-2} g^{(4)}_{\mu\nu} \) \diff x^{\mu} \diff x^{\nu}, \label{eq:feffermangraham}
  \end{align}
where $g^{(0)}_{\mu\nu} = \g_{\mu\nu}$ is the boundary metric~\eqref{eq:conformalgauge}. The subleading components of the metric are determined by (see for example ref.~\cite{Skenderis:1999nb})
  \begin{align}
  & g^{(2)}_{\mu\nu} = -\frac{1}{2} \( g^{(0)}_{\mu\nu} R^{(0)} - T_{\mu\nu} \), & g^{(4)}_{\mu\nu} = \frac{1}{4} g^{(2)}_{\mu \rho}g_{(0)}^{\rho\s}g^{(2)}_{\s\nu}, \label{eq:subleadingmetric}
  \end{align}
where terms with a $^{(0)}$ index are computed using the boundary metric $g^{(0)}_{\mu\nu}$. The Brown-York stress-energy tensor $T_{\mu\nu}$ is covariantly conserved and its trace reproduces the conformal anomaly up to a normalization,
  \begin{align}
  \nabla_{(0)}^{\mu} T_{\mu\nu} = 0, && g^{\mu\nu}_{(0)} T_{\mu\nu} = R^{(0)}. \label{eq:brownyorkeq}
  \end{align}
The solution to these equations is given by
  \begin{align}
  \begin{split}
  T_{++} &= -\frac{12}{c} L(x^+) - 2 \( \p_+\vp \p_+\vp - \p_+^2 \vp \),  \\
  T_{--} &= -\frac{12}{c} \bar L(x^-) - 2 \( \p_-\vp \p_-\vp - \p_-^2 \vp \),  \\
  T_{+-} & = -2 \p_+\p_-\vp,
    \end{split} \label{eq:conformalbrownyork}
  \end{align}
where the coefficient $c = 3l/2G$ is chosen in hindsight and $L, \bar L$ are arbitrary functions. Using $T_{\mu\nu}$ it is straightforward to compute the subleading components of the metric $g^{(2)}_{\mu\nu}$, $g^{(4)}_{\mu\nu}$ and the classical phase space of solutions consistent with the boundary conditions. A well-defined variational principle where the conformal factor $\vp$ is allowed to fluctuate requires $T_{+-} = 0$ which we interpret as the equation of motion~\eqref{eq:harmoniceq}.

These boundary conditions support a $[\VirU1]_{L}\otimes[\VirU1)]_{R}$ asymptotic symmetry group~\cite{Troessaert:2013fma} with central charge $c_{\mathrm{total}} = 0$  for the Virasoro algebras and level $k = -c/6$ for the Kac-Moody algebras. The left-moving generators of Virasoro and Kac-Moody transformations are respectively given by~\cite{Troessaert:2013fma}
  \begin{align}
  \xi[\e] = \e(x^+) \p_+, && \z[\l] = -\frac{r \l(x^+)}{2} \p_r + \frac{e^{-2\vp} \p_+ \l(x^+)}{2r^2} \p_-,\label{eq:viru1generators}
  \end{align}
while the right-moving generators are
  \begin{align}
  \bar \xi[\bar \e] = \bar \e(x^-) \p_-, && \bar \z[\bar \l] = -\frac{r \bar \l(x^-)}{2} \p_r + \frac{e^{-2\vp} \p_- \bar\l(x^-)}{2r^2} \p_-.		\label{eq:viru1generators2}
  \end{align}
The charges associated with these symmetries may be computed by standard methods~\cite{Abbott:1981ff,Iyer:1994ys,Barnich:2001jy,Barnich:2007bf}. For three-dimensional gravity the infinitesimal charge corresponding to the asymptotic Killing vector $\xi$ is given by
  \begin{align}
  \d Q_{\xi}[h, g] = \frac{1}{16 \pi G} \int_{\p\ss} \sqrt{|g|} \diff x^{\a} \ve_{\a\mu\nu} k^{\mu\nu},  \label{eq:charge}
  \end{align}
where $h_{\mu\nu} = \d g_{\mu\nu}$, $\p\ss$ is a complete spacelike surface on the boundary, and $k^{\mu\nu}$ is given by
  \begin{align}  
  k^{\mu\nu} = \xi^{\mu}\nabla^{\nu} h - \xi^{\mu} \nabla_{\rho} h^{\rho\nu} + \xi_{\rho} \nabla^{\mu} h^{\rho\nu} + \frac{1}{2} h \nabla^{\mu} \xi^{\nu} - \frac{1}{2} h_{\mu\rho} \nabla^{\rho}\xi^{\nu} + \frac{1}{2} h^{\mu\rho} \nabla^{\nu} \xi_{\rho}.  
  \end{align}

For the solution characterized by eq.~\eqref{eq:conformalbrownyork}, the charges are conserved, finite, and integrable only when $\vp$ obeys the equation of motion~\eqref{eq:harmoniceq}, in agreement with the results of ref.~\cite{Troessaert:2013fma}. The left and right-moving Kac-Moody charges are respectively given by
  \begin{align}
  U_{\l} = -\frac{1}{2\pi} \int \diff\phi \l(x^+) J(x^+), && J = k \p_+ \vp,  \label{eq:leftu1current} \\
  \overline U_{\bar \l} = -\frac{1}{2\pi} \int \diff\phi \bar\l(x^-) \bar J(x^-), && \bar J = k \p_- \vp. \label{eq:rightu1current}
  \end{align}
These currents generate two independent, centrally-extended $\U1$ algebras
  \begin{align}
  i[U_{\a},U_{\b}] =  -\frac{k\,\eta}{4\pi} \int \diff\phi \( \a'\b - \a\b' \) , &&  i[\overline U_{\bar\a},\overline U_{\bar\b}] =  -\frac{k\,\eta}{4\pi} \int \diff\phi \( \bar\a'\bar\b - \bar\a\bar\b' \) ,
  \end{align}
where $\eta = -1/2$ fixes the normalization of $\U1$ and the level is $k = -c/6$\footnote{Note that the central extension can always be set to $k = -1$ by normalization of the currents. Here we have chosen a factor of 1/2 in $\eta$ so that these Kac-Moody symmetries share the same level with the Kac-Moody symmetries found in the lightcone gauge.}. These results imply the following OPEs
  \begin{align}
  \Vev{T \{J(x^+) J(0)\}} = \frac{ k\,\eta}{(x^+)^2}, && \Vev{ T\{\bar J(x^-) \bar J(0) \}} = \frac{ k\,\eta}{(x^-)^2},  \label{eq:u1ope}
  \end{align}
where $T$ is the time-ordered product\footnote{In the remainder of this paper we adopt the standard convention where the OPE $A(x)A(0)$ is understood to hold within the expectation value of a time-ordered product of a string of operators.}. From these expressions the transformations of the currents can be deduced. Alternatively, we can obtain this result from the transformation of $\vp$ under the action of the Killing vector $\psi$,
   \begin{align}
   \d_{\psi} \vp = -e^{-2\vp} \(\L_{\psi} g\)_{+-}^{(0)}, \label{eq:conformalleading}
   \end{align}
where $\L_{\psi}$ is the Lie derivative corresponding to $\psi$. For the generators of Kac-Moody transformations we have
  \begin{align}
  \d_{\l}\vp = - \frac{\l}{2}, && \d_{\bar\l} \vp= -\frac{\bar \l}{2},  \label{eq:shiftsymmetry}
  \end{align}
which is consistent with the transformation of $J$ and $\bar J$ obtained from eq.~\eqref{eq:u1ope}.
  
Let us now consider the Virasoro charges. Using eq.~\eqref{eq:charge} we find
  \begin{align}
  Q_{\e} = -\frac{1}{2\pi} \int \diff\phi \e(x^+) [L(x^+) + T_s(x^+)], && T_s= -k (\p_+ \vp\p_+\vp - \p_+^2 \vp),  \label{eq:virasoro1}\\
  \bar Q_{\bar \e} = -\frac{1}{2\pi} \int \diff\phi \bar\e(x^-) [\bar L(x^-) + \overline T_s(x^-)], && \overline T_s = -k (\p_- \vp\p_-\vp - \p_-^2 \vp), \label{eq:virasoro2}
  \end{align}
where we recognize $T_s$ and $\overline T_s$ as the \emph{twisted} Sugawara tensors of the left and right-moving Kac-Moody algebras. That is,
  \begin{align}
  T_s = \frac{\eta^{-1}}{2k} J^2 + \p_+ J, && \overline T_s = \frac{\eta^{-1}}{2k} \bar J^2 + \p_- \bar J.
  \end{align}
In the dual theory these currents correspond to the left and right-moving components of the stress-energy tensor of the Polyakov action, while $L$ and $\bar L$ are identified with the expectation value of the of the stress-energy tensor of the matter and ghost systems. The central charge $c_M$ of the latter can be determined from the anomalous transformation of $L$ and $\bar L$ under Virasoro reparametrizations. This can be obtained from the transformation of the subleading components of the metric as in refs.~\cite{Banados:1998gg,Balasubramanian:1999re},
  \begin{align}
  \d_{\psi} g^{(2)}_{\mu\nu} = \( \L_{\psi} g \)^{(2)}_{\mu\nu}. \label{eq:conformalsubleading}
  \end{align}
Since $g^{(2)}_{\mu\nu}$ depends on $L$, $\bar L$ and $\vp$, on the left-hand side of this expression we use the transformation of $\vp$ given by eq.~\eqref{eq:conformalleading}. For the generators of Virasoro transformations the leading components transform as
  \begin{align}
  \d_{\e} \vp= \e\p_+ \vp + \frac{1}{2}\e', &&   \d_{\bar\e} \vp= \bar\e\p_- \vp + \frac{1}{2}\bar\e'. \label{eq:deltavpconformal}
  \end{align}
Then, using eq.~\eqref{eq:conformalsubleading} we find that $L$ and $\bar L$ transform anomalously
  \begin{align}
  \d_{\e} L = 2 \e' L + \e L' + \frac{c}{12} \e''', &&   \d_{\bar\e} \bar L = 2 \bar\e' \bar L + \bar \e \bar L' + \frac{c}{12} \bar\e''', \label{eq:anomalous}
  \end{align}
where $c = 3l/2G$ is the Brown-Henneaux central charge. Thus, the central charge of the matter and ghost systems in the dual theory at the boundary is $c_M = 3l/2G$.

On the other hand the central charge of the twisted Sugawara tensors, denoted here by $c_p$, is entirely determined by the Kac-Moody algebras and given by~\cite{DiFrancesco:1997ok}
  \begin{align}
  c_p = 1 + 6 k.
  \end{align}
In the semiclassical limit probed by the bulk this reduces to $c_p = -c$. We can check this result directly from the commutator of Virasoro charges
  \begin{align}
  i[Q_{\a},Q_{\b}] = Q_{\a'\b - \a\b'}, &&   i[\bar Q_{\bar\a},\bar Q_{\bar\b}] = \bar Q_{\bar\a'\bar\b - \bar\a\bar\b'},  \label{eq:virasorocommutators}
  \end{align}
which confirms that the total central charge $c_{\mathrm{total}} = c_M + c_p$ vanishes. These expressions can also be used to check that $T_s$ and $\overline T_s$ transform as tensors of weight (2,0) and (0,2), respectively, as expected from the current algebra\footnote{Here we have assumed that $L$ and $\bar L$ commute with the Kac-Moody currents; this is confirmed below.},
  \begin{align}
  T_s(x^+) T_s(0) &= \frac{- c/2}{\( x^+\)^4} + \frac{2T_{s}(0)}{\( x^+ \)^2} + \frac{\p_+ T_{s}(0)}{x^+ }, \label{eq:sugawaraope}\\
  \overline T_s(x^-) \overline T_s(0) &= \frac{-c/2}{\( x^-\)^4} + \frac{2\overline T_{s}(0)}{\( x^- \)^2} + \frac{\p_- \overline T_{s}(0)}{x^-}. \label{eq:sugawaraope2}
  \end{align}

Alternatively, we can find the central charge of the matter and ghost system by computing the commutators associated with the Killing vectors~\cite{Troessaert:2013fma}
  \begin{align}
  \tilde\xi[\e] = \xi[\e] + \z[\e'], &&  \tilde{\bar\xi}[\bar\e] = \bar\xi[\bar\e] + \bar\z[\bar\e'], \label{eq:untwistedkilling}
  \end{align}
which are none other than the asymptotic Killing vectors of AdS$_3$ with Brown-Henneaux boundary conditions. For these Killing vectors the commutators of charges are given by eq.~\eqref{eq:virasorocommutators} with central extension $\tilde c_{\mathrm{total}} = 3l/2G$. This is a consequence of the fact that $\tilde\xi$ and $\tilde{\bar\xi}$ untwist the Sugawara tensors in eqs.~\eqref{eq:virasoro1} and~\eqref{eq:virasoro2}. Hence $c_p = 1$ and in the semiclassical limit $\tilde c_{\mathrm{total}} = c_M = 3l/2G$.

Finally let us consider the commutators between Virasoro and Kac-Moody charges. These are given by
  \begin{align}
  i[Q_{\e}, U_{\l}] = -U_{\e\l'} - \frac{k}{4\pi} \int \diff\phi\, \e \l'', &&   i[\bar Q_{\bar\e}, \overline U_{\bar\l}] = -\overline U_{\bar\e\bar\l'} - \frac{k}{4\pi} \int \diff\phi\, \bar\e \bar\l''
  \end{align}
and agree with the commutators given in ref.~\cite{Troessaert:2013fma}. Using the expressions for the Virasoro and Kac-Moody charges it is easy to check that these commutators follow from the OPEs
  \begin{align}
  \begin{split}
  T_{s}(x^+)J(0) = \frac{k}{\(x^+\)^3} + \frac{J(0)}{\( x^+ \)^2} + \frac{\p_+ J(0)}{x^+}, \\
  \overline T_{s}(x^-)\bar J(0) = \frac{k}{\(x^-\)^3} + \frac{\bar J(0)}{\( x^- \)^2} + \frac{\p_- \bar J(0)}{x^-},
  \end{split}  \label{eq:viru1ope}
  \end{align}
which imply that $L, \bar L$ commute with the generators of Kac-Moody transformations. This result can also be checked by inspection of the subleading components of the metric. Using eq.~\eqref{eq:conformalsubleading} and the Kac-Moody Killing vectors $\z[\l]$~\eqref{eq:viru1generators} and $ \bar\z[\bar\l]$~\eqref{eq:viru1generators2} we find that
  \begin{align} 
  \d_{\l} L = 0, && \d_{\bar \l} \bar L = 0.
  \end{align}

  We have thus identified the currents generating the Virasoro and Kac-Moody transformations and showed that the OPEs deduced from the commutator of charges yield a consistent algebra. %We have thus identified the currents generating the Kac-Moody transformations and showed that their OPEs are consistent with their treatment as operators of weight (1,0) ($J$) and (0,1) ($\bar J$) in a field theory. We have also found the currents generating Virasoro transformations. These consist of functions $L$ and $\bar L$ that transform anomalously as operators of weight (2,0) and (0,2) in a field theory~\eqref{eq:anomalous}. The same conclusion is reached for the twisted Sugawara tensors $T_s$, $\overline T_s$ whose OPEs are determined by the Kac-Moody algebra as expected. 
In the next section we show that these currents match those of the dual theory at the boundary.

%%%%%%%%%%%%%%%%%%%%%%%%%%%%%%%%%%%%%%%%%%%%%%%%%%%
\subsection{Dual theory of gravity}
%%%%%%%%%%%%%%%%%%%%%%%%%%%%%%%%%%%%%%%%%%%%%%%%%%%
\label{se:conformaldualtheory}

Let us now consider the dual theory which is described by 
  \begin{align}
  \Z = \int \D\phi_a \D\vp\, \mathrm{exp}\,i\lB  S_{m+g}[\phi_a] + \frac{k}{16\pi} S_{p}[\vp]\rB, 
  \end{align}
where $k = -c_M/6$ and $c_M$ is the central charge of the matter and ghost systems described by $S_{m+g}$. This action is assumed to be conformally invariant at the classical level but since $c_M \ne 0$ the measures of the matter and ghost fields $\phi_a$ contribute to the quantum anomaly. The Polyakov action can be more conveniently written as
  \begin{align}
  S_p = 2 \int \sqrt{|\g|} \lB \frac{1}{2} \p_{\mu} \psi \p^{\mu} \psi - R \psi \rB, \label{eq:localpolyakov}
  \end{align}
where we must substitute the expression for $\psi$ given by $\Box \psi = - R$. In other words, eq.~\eqref{eq:localpolyakov} reproduces the Polyakov action when computed on-shell. For the conformal gauge we have $\psi = 2 \vp$ and the Polyakov action reduces to the action of a timelike scalar field\footnote{Recall that in our conventions the boundary metric reads $\diff s_b^2 = e^{2\vp} \( -\diff t^2 + \diff x^2\)$.}. For more general reference metrics neither the conformal coupling nor the cosmological constant term vanish. 

It is easy to show that the stress-energy tensor is given by
  \begin{align}
  T_{++} &= t_{++} - k \( \p_+ \vp \p_+ \vp - \p_+^2\vp \), \label{eq:dualtmunu1} \\
  T_{--} &= t_{--} - k\( \p_- \vp \p_- \vp - \p_-^2\vp \), \label{eq:dualtmunu2} \\  
  T_{+-} &= -k \p_+\p_-\vp,
  \end{align}
where $t_{\mu\nu}$ is the stress-energy tensor of the matter and ghost fields. This is in perfect agreement with the Brown-York stress-energy tensor given in eq.~\eqref{eq:conformalbrownyork} after normalization by a factor of $k/2$. Thus $L$ and $\bar L$ are to be identified with the expectation values of the $t_{++}$ and $t_{--}$ components of the stress-energy tensor of the matter and ghost systems. In particular, their central charge can be determined in the bulk from the anomalous transformation of $L$ and $\bar L$ and is given by $c_M = 3l/2G$ in the semiclassical approximation.
 
Since the Polyakov action reduces to that of a free scalar field, the action is shift-invariant. When the cosmological constant does not vanish this shift symmetry is still present when accompanied by a Weyl rescaling of the reference metric. Indeed, one can think of $\vp$ as a St\"uckelberg field that restores the Weyl invariance of the matter plus ghost theory, despite the fact that $c_M$ does not vanish. The Noether currents corresponding to the shift in $\vp$ parametrized by eq.~\eqref{eq:shiftsymmetry} are given by
  \begin{align}
  J^- &= k \p_+ \vp, && & J^+& = 0, \\
  \bar J^-& = 0, && &\bar J^+ &= k \p_- \vp,
  \end{align}
where the upper indices denote spacetime indices and we have assumed that the matter plus ghost action is Weyl invariant. These currents match the $\U1$ Kac-Moody currents found in the bulk, eqs.~\eqref{eq:leftu1current},~\eqref{eq:rightu1current}. Furthermore, it is not difficult to show that the OPEs given in eq.~\eqref{eq:u1ope} and deduced from the commutators of Kac-Moody charges, agree with the OPEs obtained in the dual theory from quantization of $\vp$, whose OPE is given by
  \begin{align}
  \vp(x)\vp(x') = -\frac{1}{2k} \log |x^+ - x'^+||x^- - x'^-|.  \label{eq:conformaldofope}
  \end{align}

On the other hand the currents generating the Virasoro transformations -- where $\vp$ transforms with an additional shift as in eq.~\eqref{eq:deltavpconformal} -- correspond to the $T_{++}$ and $T_{--}$ components of the stress-energy tensor. The contribution from the Polyakov side of the action is the twisted Sugawara tensor, whose central charge is given by
  \begin{align}
  c_p = 1 + 6k.
  \end{align}
In the quantum theory the constant $k$ in front of the Polyakov action is renormalized in such a manner that $c_M + c_p = 0$. In the semiclassical limit where $k$ is large we have $c_p = 6k = -c_M$ in agreement with the bulk computation. The other OPEs between the Sugawara tensors and the Kac-Moody currents, eqs.~\eqref{eq:sugawaraope},~\eqref{eq:sugawaraope2}, and~\eqref{eq:viru1ope}, can also be seen to follow from the OPE~\eqref{eq:conformaldofope}.

It is important to note that for the boundary conditions studied in this section, which are conformally flat $\diff s_b^2 = e^{2\vp} \g^{(0)}_{\mu\nu} \diff x^{\mu}\diff x^{\nu}$, we have $R = -2\Box^{(0)}\vp$ so the Polyakov action is simply that of a free scalar. For more general boundary metrics the term $R \psi$ receives contributions proportional to $R^{(0)} \vp$, where $R^{(0)}$ is the Ricci scalar of the boundary metric $\g^{(0)}_{\mu\nu}$. This improvement term is responsible for the twisting of the stress-energy tensor. Hence, when the topology of the boundary is that of the cylinder or the torus and $R^{(0)} = 0$, one could argue that there is no improvement term and that the stress-energy tensor is not twisted. This would be consistent with the bulk theory since, as we have seen, there is a linear combination of Killing generators that lead to an untwisted Sugawara tensor~\eqref{eq:untwistedkilling}. However, if we want to define the dual CFT on surfaces with higher genus, a twisted stress-energy tensor for which the total central charge vanishes is unavoidable. Since we are interested in the interpretation of free boundary conditions in the context of the AdS/CFT correspondence, we have assumed that the improvement term contributes to the stress-energy tensor regardless of the topology of the boundary.

Let us now consider the implications of the equations of motion~\eqref{eq:boundarytmunu}. The vanishing of the $T_{+-}$ component of the stress-energy tensor corresponds to the equation of motion for $\vp$. The fact that $c_M + c_p = 0$ then implies that ${T_{+-}}$ does not receive corrections at the quantum level. Indeed, we could have obtained the same results of this section if instead of adding the Polyakov action we had studied the Ward identities of the matter plus ghost systems which are sensitive to contributions from the measure. The other equations of motion of the theory prior to gauge fixing,
  \begin{align}
  {T_{++}} = 0, && {T_{--}} = 0,
  \end{align}
are interpreted as constraints on the physical states of the theory. In particular they require that physical states are singlets of the Virasoro algebra. Hence the weight of operators associated with Virasoro transformations must vanish. On the other hand the weights associated with the $\U1$ Kac-Moody transformations, i.e. Weyl scalings, receive gravitational corrections. In particular, the gravitationally-dressed weights of scalar primary fields are given by~\cite{David:1988hj,Distler:1988jt}
  \begin{align}
  \l = h + \frac{\l(1-\l)}{k+2}, \label{eq:dressing}
  \end{align}
where $h$ is the weight in the absence of gravity. In the next section we show that this result is reproduced by the bulk theory in the semiclassical limit.

Note that these constrains may play an important role in the unitarity of the theory because, as we have seen, the level of the Kac-Moody algebra is negative. This is consistent with the fact that the conformal factor is timelike and hence, in the absence of constraints, the theory is not unitary. However, as in the quantization of the critical string, the constraints $T_{++} = T_{--} = 0$ may be used to remove the negative-norm states from the Hilbert space of the theory.

%%%%%%%%%%%%%%%%%%%%%%%%%%%%%%%%%%%%%%%%%%%%%%%%%%%
\subsection{Virasoro and \texorpdfstring{$\U1$}{U1} weights for scalar operators}
%%%%%%%%%%%%%%%%%%%%%%%%%%%%%%%%%%%%%%%%%%%%%%%%%%%
\label{se:weights}

Let us consider a free scalar field of mass $m$ in the bulk and study its asymptotic behavior. In AdS$_3$ with Brown-Henneaux boundary conditions we have
  \begin{align}
  \phi(r,x) \sim r^{-\dd_-} \a(x) + r^{-\dd_+} \b(x), \label{eq:bulkscalar}
  \end{align}
where we have assumed $\dd_+ \ne \dd_-$ and $\dd_{\pm}$ are the roots of (see for example~\cite{Aharony:1999ti})
  \begin{align}
  \dd(\dd - 2) = m^2.
  \end{align}
The non-normalizable mode $\a(x)$ is interpreted as a source for the operator $\O(x)$ in the dual theory while $\b(x)$ is proportional to its expectation value. The weight of this operator can be deduced from the transformation of $\phi$ under the Virasoro symmetry. For AdS$_3$ with Brown-Henneaux boundary conditions the left-moving generator of Virasoro transformations is given by
  \begin{align}
  \xi_{\mathrm{BH}}[\e] = \e(x^+) \p_+ - \frac{r\e'(x^+)}{2}\p_r + \frac{\e''(x^+)}{2 r^2}\p_-.  \label{eq:killing_vir_bh}
  \end{align}
Under the action of $\xi_{\mathrm{BH}}$ the scalar field transforms as
  \begin{align}
  \L_{\xi_{\mathrm{BH}}} \phi \sim r^{\dd_-} \( \frac{\dd_-}{2}\e'\a +\e\, \p_+\a \) + r^{\dd_+} \( \frac{\dd_+}{2}\e'\b +\e\,\p_+ \b \),
  \end{align}
from which we deduce that the dual operator $\O$ has left (and right) weights $h  = \bar h = \dd_+/2$.

It is not surprising that a massive scalar field in AdS$_3$ with boundary conditions given in eq.~\eqref{eq:bc1} obeys the same asymptotic behavior as eq.~\eqref{eq:bulkscalar}. Unlike the asymptotic Killing vector $\xi_{\mathrm{BH}}$, however, the generators of Virasoro transformations in eqs.~\eqref{eq:viru1generators}, and~\eqref{eq:viru1generators2} have vanishing $\xi^r$ components. Thus, scalar operators in the dual theory have a vanishing weight under Virasoro transformations,
  \begin{align}
  \L_{\e} \phi \sim r^{\dd_-} \( \e\, \p_+\a \) + r^{\dd_+} \( \e\, \p_+\b\).
  \end{align}
This result is to be expected since we must enforce the constraints $T_{++} = 0$ and $T_{--} = 0$ in the dual theory, and a vanishing Virasoro weight means that all scalar operators are singlets of the Virasoro symmetry. On the other hand the generators of Kac-Moody transformations in eqs.~\eqref{eq:viru1generators} and~\eqref{eq:viru1generators2} do have non-vanishing $\xi^r$ components. Under the left-moving $\U1$ transformation parametrized by $\s(x^+)$ the scalar field transforms as
  \begin{align}
  \L_{\l} \phi \sim r^{\dd_-} \(  \frac{\dd_-}{2}\s\a \) + r^{\dd_+} \(  \frac{\dd_+}{2}\s\b \),
  \end{align}
and similarly for the right-moving $\U1$. Thus the gravitationally-dressed weights are given by $\l = \bar \l = \dd_+/2$ which corresponds to the semiclassical limit of eq.~\eqref{eq:dressing}.

%%%%%%%%%%%%%%%%%%%%%%%%%%%%%%%%%%%%%%%%%%%%%%%%%%%
\section{Boundary conditions with \texorpdfstring{$\SL2R$}{SL(2,R)} symmetry}
%%%%%%%%%%%%%%%%%%%%%%%%%%%%%%%%%%%%%%%%%%%%%%%%%%%
\label{se:lightconegauge}

It would be surprising if the conclusions obtained in Section~\ref{se:conformalgauge} did not hold in other gauges. In this Section we perform a similar analysis in a generalization of the lightcone gauge that can be obtained by performing the change of coordinates $x^- \ra e^{nx^-}$, $n \in \mathbb{Z}$ in eq.~\eqref{eq:lightconegauge}. The boundary metric reads,
  \begin{align}
  \mathrm{ds}_{b}^{2} = -e^{nx^-} \diff x^{+} \lb \diff x^{-} + \mu(x) \diff x^{+} \rb, \label{eq:newlightconegauge}
  \end{align}
and will be referred to as the lightcone gauge for the remainder of this paper. We will also consider a slight modification of the boundary conditions of ref.~\cite{Avery:2013dja} where the $g^{(2)}_{--}$ component of the metric is arbitrary, 
  \begin{align}
  & g_{rr} = r^{-2} + \O(r^{-4}),  & g_{+-} &= -e^{nx^-} r^2/2 + \O(r^0),    \notag      \\
  & g_{r+} = \O(r^{-1}),           & g_{++} &= - e^{nx^-} r^2 \mu(x) + \O(r^0),   \label{eq:lightconebc}     \\
  & g_{r-} = \O(r^{-3}),           & g_{--} &= \O(r^{0}). \notag
  \end{align}
A well-defined variational principle where $\mu$ is allowed to vary then requires that the $T_{--} \propto g^{(2)}_{--}$ component of the Brown-York stress-energy tensor vanishes. Thus we must either add improvement terms to the action so that $\mu$ is allowed to fluctuate, or set $T_{--} = 0$ as an equation of motion. Since we would like to interpret AdS$_3$ with these boundary conditions as a theory of gravity at the boundary we choose the second option.

Unlike the original boundary conditions of ref.~\cite{Avery:2013dja}, the boundary conditions given above do not admit compactification of the $\phi$ coordinate. Hence in the limit $\mu \ra 0$ the dual theory at the boundary is defined on the plane. It is possible to analytically continue $n \ra in$ to study the theory on the cylinder but then the bulk and boundary metrics are complex. This is not too problematic in the dual theory since the conformal factor drops out of the matter, ghost, and Polyakov actions, and it is possible to restrict the theory to the real plane. In the remainder of this paper we take $n \in \mathbb{Z}$ but our conclusions hold for complex $n$.

The most general solution obeying these boundary conditions may be written once again using Fefferman-Graham coordinates~\eqref{eq:feffermangraham}. First let us note that it is convenient to regard the lightcone parameter $\mu$ as the coefficient of a Beltrami differential~\cite{Polyakov:1987zb}. Indeed, the lightcone gauge may be obtained from the conformal gauge by the change of coordinates $x^+ \ra x^+$, $x^- \ra f(x^-)$, and $2\vp \ra - \log(\p_- f) + n x^-$, whereby $\mu$ is given by
  \begin{align}
  \mu = \frac{\p_+ f}{\p_- f}. \label{eq:beltrami}
  \end{align}
It is then easy to see that $\mu$ has weight $(1,-1)$. The Brown-York stress-energy tensor satisfying eq.~\eqref{eq:brownyorkeq} is given by
  \begin{align}
  T_{++} &=  -\frac{12}{c} L(x^+) - (\p_+ -2 \mu\p_-) \( \p_- \mu + n \mu \)  + \( \mu \p_-^2 \mu - \frac{1}{2} \p_- \mu \p_- \mu  - \frac{n^2}{2} \mu^2 \) + \mu^2 T_{--} , \notag \\
  T_{--} & = - \frac{12}{c}\bar L(x^-) - \( \{f,x^-\} + \frac{n^2}{2} \),\label{eq:lightconebrownyork} \\
  T_{+-} & = \p_-^2 \mu + n \p_-\mu + \mu T_{--},  \notag
  \end{align}
where $c = 3l/2G$, $\{f,x^-\}$ is the Schwarzian derivative of $f$ with respect to $x^-$, and 
$\bar L(x^-)$ is a function obeying
  \begin{align}
  \( \p_+ - \mu \p_- \) \bar L - 2 \bar L \p_- \mu = 0. \label{eq:lightconecondition}
  \end{align}
The equation of motion $T_{--} = 0$ and the identity $\p_-^3 \mu \equiv  \( \p_+ - \mu \p_- \) \{f,x^-\} - 2  \{f,x^-\} \p_- \mu$ then yield the following equation for $\mu$, 
  \begin{align}
  \p_-^3 \mu - n^2 \mu = 0. \label{eq:mueom}
  \end{align}
We identify $L(x^+)$ with the expectation value of the generator of Virasoro transformations of the matter and ghost systems, and $\bar L(x)$ with the expectation value of the $t_{--}$ component of their stress-energy tensor. In the next section we will see that condition~\eqref{eq:lightconecondition} corresponds to covariant conservation of $t_{\mu\nu}$.

The boundary conditions given in eq.~\eqref{eq:lightconebc} support a $\VirSL2R$ asymptotic symmetry group~\cite{Avery:2013dja} with total central charge $c_{total} = 0$ and Kac-Moody level $k = -c/6$. The asymptotic Killing vectors generating the Virasoro and Kac-Moody symmetries are respectively given by
  \begin{align}
  & \xi[\e] = \e(x^+) \p_+ - \frac{1}{n}\e'(x^+) \p_-, \label{eq:killing_vir} \\
  & \z[\ll] = \frac{e^{-nx^-}\p_- \ss[\ll]}{2r^2}   \p_+ - \frac{r\ss[\ll]}{2}  \p_r + \ll \p_{-} + \( \frac{e^{-nx^-}\( \p_+ - 2\mu\p_- \) \ss[\ll] }{2r^2}\p_-\)_{\bullet} \label{eq:killing_sl2r},
  \end{align}
where $(\quad)_{\bullet}$ is a pure gauge term that will be useful later. Here $\ss[\ll]$ is given by
  \begin{align}
  \ss[\ll] = \p_- \ll + n \ll,
  \end{align}
and the parameter $\ll(x)$ obeys the same equation as $\mu$, i.e.
  \begin{align}
  \p_-^3 \ll - n^2 \p_-\ll = 0. \label{eq:lambdaeom}
  \end{align}
The Killing vectors for the case $n = 0$ can be obtained by performing a change of coordinates $e^{nx^-} \ra x^-$ where $\ll$ and $\mu$ transform as tensors of weight $(1,-1)$ (see eq.~\eqref{eq:beltrami}). This is only necessary when the $n \ra 0$ limit is divergent as in the Virasoro Killing vector $\xi[\e]$. In expressions without factors of $1/n$ setting $n = 0$ is equivalent to the aforementioned change of coordinates.

The charges of the theory are conserved, finite, and integrable only when $T_{--} = 0$. Using eq.~\eqref{eq:charge} the charge associated with the Kac-Moody generator $\z[\ll]$ reads 
  \begin{align}
  & U[\ll] = -\frac{1}{2\pi}\int \diff \phi\, \frac{k}{2} \( \ll\p_-^2\mu - \p_-\ll\p_-\mu +\p_-^2\ll \mu - n^2\ll\mu \), \label{eq:sl2rcharge}
   \end{align}
where we recall that the coordinate $\phi$ is not compact. Equation~\eqref{eq:lambdaeom} allows us to parametrize $\ll$ by
  \begin{align}
  \ll & = \frac{1}{n} \lb \l_0(x^+) + \l_{+}(x^{+})e^{nx^-} + \l_{-}(x^{+}) e^{-nx^-} \rb. \label{eq:sl2rparameter}
  \end{align}
Then the Kac-Moody charges can be written as 
  \begin{align}
  U_{a}[\l_a] = -\frac{1}{2\pi} \int \diff \phi\,\l_a(x^+) J_a(x^+), \quad\quad a=\{0,\pm\}, \label{eq:sl2rcharge2}
  \end{align}
where the left-moving currents $J_a$ are given by
  \begin{align}
  J_0 = \frac{k}{2n} \( \p_-^2 \mu - n^2 \mu \), && J_{\pm} = \frac{k}{2n} e^{\pm nx^-}\( \p_-^2\mu \mp n\p_-\mu  \). \label{eq:sl2rcurrents}
  \end{align}
We stress that the $n = 0$ case is well defined and can be obtained by setting $n = 0$ in eq.~\eqref{eq:sl2rcharge}  and performing a change of coordinates $e^{nx^-} \ra x^-$ in eq.~\eqref{eq:sl2rparameter}. Alternatively, one can change the coordinates directly in the expression for the currents~\eqref{eq:sl2rcurrents}. 

In Section~\ref{se:lightconecft} we will show that $J_a$ match the $\SL2R$ currents of the dual theory at the boundary. That these currents generate an $\sl2r$ Kac-Moody algebra with level $k = -c/6$ can be seen from the commutator of charges,
   \begin{align}
   i [U_a[\a],U_b[\b]] = \ve_{abc} \eta^{cd} U_{d}[\a\b] - \frac{k\,\eta_{ab}}{4\pi} \int \diff\phi \(\a' \b - \a\b'\), \label{eq:sl2rcomm}
   \end{align}
where $\ve_{0+-} = -1$ and the only non-vanishing components of $\eta_{ab}$ are $\eta_{00} = -1/2$ and $\eta_{+-} = \eta_{-+} = 1$. Using eq.~\eqref{eq:sl2rcharge2} we can recover the transformation of the currents leading to the commutator~\eqref{eq:sl2rcomm}. These can be expressed more conveniently as an OPE
  \begin{align}
  J_a(x^+) J_b(0) = \frac{k\,\eta_{ab}}{(x^+)^2} + \frac{\ve_{abc}\eta^{cd} J_d (x^+)}{x^+}, \label{eq:sl2rOPE}
  \end{align}
which agrees with the expression found in ref.~\cite{Alekseev:1988ce} and more readily demonstrates that we have an $\SL2R$ algebra. An alternative way of deriving this result is to consider the transformation of the leading components of the metric as in Section~\ref{se:conformalgauge}. The  solution for $\mu$ that is consistent with the expression for the currents in eq.~\eqref{eq:sl2rcurrents} is given by
    \begin{align}
    \mu = \frac{1}{kn} \(  -2J_0 + J_- e^{nx^-} + J_+ e^{-nx-}  \). \label{eq:musolution}
    \end{align}
Then the transformation of the currents under the action of the Killing vector $\psi$ can be obtained from
   \begin{align}
   \d_{\psi} \mu = -\(\L_{\psi} g\)_{++}^{(0)}, \label{eq:leading}
   \end{align}
where $\L_{\psi}$ is the Lie derivative corresponding to $\psi$. For the Kac-Moody transformations generated by~eq.\eqref{eq:killing_sl2r} we obtain
  \begin{align}
  \d_{\ll} \mu = \( \p_+ -\mu\p_- \)\ll + \ll \p_- \mu, \label{eq:deltamulightcone}
  \end{align}
and it is not difficult to check that the transformations of the currents deduced from this equation are consistent with the commutator~\eqref{eq:sl2rcomm} and the OPE~\eqref{eq:sl2rOPE}.

Let us now consider the charge associated with the Virasoro symmetry. Using the asymptotic Killing vector~\eqref{eq:killing_vir} we find
  \begin{align}
  Q_{\e} = -\frac{1}{2\pi} \int \diff\phi\,\e(x^+) \lb L(x^+) + T_s(x^+) \rb, \label{eq:virasorocharge}
  \end{align}
where $T_s$ is the \emph{twisted} Sugawara tensor given by
  \begin{align}
  T_s & = \frac{k}{2} \( \mu\p_-^2\mu - \frac{1}{2} \p_-\mu\p_-\mu - \frac{n^2}{2} \mu^2  \) + \frac{k}{2n} \p_+ \( \p_-^2 \mu - n^2 \mu \) \label{eq:twisted1}.
  \end{align}
In terms of the $\SL2R$ currents the twisted Sugawara tensor can be more conveniently written as
  \begin{align}
  T_s & = \frac{1}{2k} \eta^{ab} J_a J_b + \p_+ J_0. \label{eq:twisted2}
  \end{align}
We must note that this expression has been obtained in the semiclassical limit where $k$ is large. Hence it is not surprising that the first term in eq.~\eqref{eq:twisted2} differs from the standard form of the Sugawara tensor of $\SL2R$ where $k \ra k + 2$.

As mentioned earlier we identify $L$ with the expectation value of the generator of Virasoro transformations of the matter plus ghost part of the dual theory. Then the central charge of the matter and ghost systems can be determined from the anomalous transformation of $L$ under Virasoro reparametrizations. The latter can be obtained from the change in the subleading components of the metric using eq.~\eqref{eq:conformalsubleading}. Since $g^{(2)}_{\mu\nu}$ depends on $L$ and $\mu$, on the left-hand side of this expression we use the transformation of $\mu$ given by eq.~\eqref{eq:leading}. For the Virasoro generator~\eqref{eq:killing_vir} $\mu$ transforms as
  \begin{align}
  \d_{\e} \mu = \e \p_+ \mu + \p_+\e \mu -\frac{1}{n} \(\p_+^2 \e + \p_+\e\p_-\mu \). \label{eq:muundervir}
  \end{align}
Then, using eq.~\eqref{eq:conformalsubleading} we find that $L$ transforms as
  \begin{align}
  \d_{\e} L = 2 \e' L + \e L' + \frac{c}{12} \e''',
  \end{align}
where $c = 3l/2G$ is the Brown-Henneaux central charge. We have thus obtained the same result found in the boundary conditions for the conformal gauge: the central charge of the matter and ghost systems is $c_M =  3l/2G$.

On the other hand, the central charge of the twisted Sugawara tensor is determined by the $\SL2R$ algebra and is given by~\cite{DiFrancesco:1997ok}
  \begin{align}
  c_p = \frac{3 k}{k+2} + 6k.
  \end{align}
In the semiclassical limit where our analysis is valid this reduces to $c_p =  -c$. This is consistent with the commutator of charges since the total central charge, $c_{total} = c_M + c_p$, vanishes,
  \begin{align}
  i[Q_{\e}, Q_{\s}] = Q_{\e'\s - \e\s'}. \label{eq:lightconevircommutator}
  \end{align}
As in Section~\ref{se:conformalgauge}, eqs.~\eqref{eq:lightconevircommutator} and~\eqref{eq:virasorocharge} can be used to determine the $T_s(x^+)T_s(0)$ OPE which is consistent with the current algebra given in eq.~\eqref{eq:sl2rOPE} in the semiclassical approximation. Mirroring the discussion of Section~\ref{se:conformalgauge} it is also possible to recover the central charge of the matter and ghost fields by untwisting the Sugawara tensor in eq.~\eqref{eq:virasorocharge}. This is accomplished by the Killing generator~\cite{Avery:2013dja}
  \begin{align}
  \tilde\xi[\e] = \xi[\e]+\frac{1}{n}\z[\e'].
  \end{align}
Then in the semiclassical limit the central charge of the Sugawara tensor is $c_p \sim \O(1)$ and the commutator of charges~\eqref{eq:lightconevircommutator} picks up a central charge equal to $c_{total} = c_M = 3l/2G$.

The last ingredient in the $\VirSL2R$ asymptotic symmetry algebra consists of the commutators between Virasoro and Kac-Moody charges which are given by
  \begin{align}
  i[Q_{\e}, U_0[\l]] &= -U_0 [\e \l'] - \frac{k}{4\pi} \int \diff \phi \e \l'', \notag \\
  i[Q_{\e}, U_+[\l]] &= -U_+[\e\l' -\e'\l], \label{eq:commvirsl2r} \\
  i[Q_{\e}, U_-[\l]] &= -U_-[\e\l' + \e'\l]. \notag
  \end{align}
Using eq.~\eqref{eq:sl2rcharge2} it is not difficult to check that these expressions are consistent with the transformation of the currents obtained from eq.~\eqref{eq:muundervir}. One can also show that these transformations follow from the OPEs
  \begin{align}
  T_s(x^+) J_0(0) &= \frac{k}{\(x^+\)^3} + \frac{J_0}{\( x^+ \)^2} + \frac{\p_+ J_0}{x^+},\\
  T_s(x^+) J_+(0)  &= \frac{2J_+}{\( x^+ \)^2} + \frac{\p_+ J_+}{x^+},\\
  T_s(x^+) J_- (0)  &= \frac{\p_+ J_-}{x^-}, \label{eq:virsl2rOPE3}
  \end{align}
which in the large-$k$ limit are a direct consequence of the $J_a J_b$ OPE given in eq.~\eqref{eq:sl2rOPE}. In particular, this implies that $L$ does not transform under the Kac-Moody symmetry. As in Section~\ref{se:conformalgauge} this result can be checked from the transformation of the subleading components of the metric using eq.~\eqref{eq:conformalsubleading}. Note, however, that the right-hand side of eq.~\eqref{eq:conformalsubleading} is sensitive to subleading terms in the the asymptotic Killing vectors. These terms correspond to pure gauge transformations since their charges vanish, but we find that they are necessary for consistency of the transformations at subleading order. This is the reason why a pure gauge term denoted by $(\quad)_{\bullet}$ has been added to eq.~\eqref{eq:killing_sl2r}. Otherwise we would find that $\d_{\ll} L = F(x^+,x^-)$ for some function $F$, which is clearly inconsistent since the $\VirSL2R$ algebra is left-moving. For the $\SL2R$ generator given in eq.~\eqref{eq:killing_sl2r} we find
  \begin{align}
  \d_{\ll} L = 0,
  \end{align}
which is consistent with the results obtained above.

The fact that we have obtained a twisted Sugawara tensor in the generator of Virasoro transformations is consistent with the constraints on the dual theory at the boundary. There we must impose $T_{+-} = 0$ on physical states and it is readily seen that $T_{+-}$ in eq.~\eqref{eq:lightconebrownyork} is proportional to $J_-$. Thus, it follows from the constraint that $J_-$ is an operator of weight zero under Virasoro transformations which is precisely what eq.~\eqref{eq:virsl2rOPE3} is telling us. Similarly, the constraint $T_{++} = 0$ implies that $L + T_s$ must vanish on physical states so that physical states in the dual theory are singlets of the Virasoro symmetry.

%%%%%%%%%%%%%%%%%%%%%%%%%%%%%%%%%%%%%%%%%%%%%%%%%%%
\section{Conformal field theory in the lightcone gauge}
%%%%%%%%%%%%%%%%%%%%%%%%%%%%%%%%%%%%%%%%%%%%%%%%%%%
\label{se:lightconecft}

Let us now turn to the dual theory at the boundary described by 
  \begin{align}
  \Z = \int \D\phi_a \D\mu\, \mathrm{exp}\,i\lB  S_{m+g}[\phi_a,\mu] + \frac{k}{16\pi} S_{p}[\mu]\rB,
  \end{align}
where $k = -c_M/6$ and $c_M$ is the central charge of the matter and ghost systems described by $S_{m+g}$. As in Section~\ref{se:conformaldualtheory}, the Polyakov action can be more conveniently written as
  \begin{align}
  S_p = 2 \int \sqrt{|\g|}\diff^2x \lB \frac{1}{2} \p_{\mu} \psi \p^{\mu} \psi - R \psi \rB,
  \end{align}
where we must substitute the expression for $\psi$ given by $ \Box \psi = - R$. For the lightcone gauge~\eqref{eq:newlightconegauge} we have $\psi = -\log(\p_- f) + n x^-$ from which it follows that 
  \begin{align}
  S_p = 2 \int \diff^2x \lB\frac{\p_-^2 f}{\p_- f} \p_- \mu - n^2 \mu\rB. \label{eq:noncovariantpolyakov}
  \end{align}
The stress-energy tensor is given by
  \begin{align}
  T_{++} &= t_{++} - \mu^2 t_{--} - \frac{k}{2} \lb (\p_+ - 2\mu\p_-)(\p_-\mu+n\mu) - \(\mu\p_-^2 \mu - \frac{1}{2}\p_-\mu\p_-\mu-\frac{n^2}{2}\mu^2 \)\rb, \notag \\
  T_{+-} &= t_{+-} - \mu t_{--} + \frac{k}{2}\( \p_-^2 \mu+ n \p_-\mu \), \label{eq:lightconetmunu}\\
  T_{--} &= t_{--} - \frac{k}{2} \( \{f,x^-\} +\frac{n^2}{2} \) \notag
  \end{align}
where we have used the equations of motion $T_{--} = 0$ and $t_{\mu\nu}$ is the stress-energy tensor of the matter and ghost systems. Eq.~\eqref{eq:lightconetmunu} agrees with the Brown-York stress-energy tensor~\eqref{eq:lightconebrownyork}, after normalization by a factor of $k/2$, provided that (1) $t_{+-} -\mu t_{--} = 0$, (2) $\bar L = \Vevm{t_{--}}$ and (3) $L = \Vevm{t_{++} - \mu^2 t_{--}}$\footnote{Here $\Vevm{\dots}$ denotes the expectation value on a state of the
matter and ghost system.}. 

The first condition follows from conformal invariance of the matter and ghost systems. In order to see this consider writing the boundary metric~\eqref{eq:newlightconegauge} as
  \begin{align}
  \g_{\mu\nu} = e^{n x^-}\( \eta_{\mu\nu} - \d_{\mu+}\d_{\nu+} \mu \),
  \end{align}
where $\eta_{\mu\nu}$ is the metric of flat space. Then the lightcone gauge may be thought as a perturbation of a conformal field theory by a marginal operator of weight (1,1)
  \begin{align}
  S[\phi_a,\g] = S[\phi_a,\eta] + \int \sqrt{|\eta|}\diff^2x\,\mu t_{--}  \label{eq:perturbation}
  \end{align}
Due to this perturbation the hitherto vanishing $t_{+-}$ component of the stress-energy tensor receives a contribution equal to $\mu t_{--}$ so that condition (1) is satisfied. 

The second condition constraints $t_{--}$ to obey~\eqref{eq:lightconecondition}
  \begin{align}
  (\p_+ - \mu \p_-) t_{--} - 2 t_{--} \p_-\mu = 0. \label{eq:eomconstraint}
  \end{align}
Since $t_{+-} = \mu t_{--}$, eq.~\eqref{eq:eomconstraint} follows from covariant conservation of the stress-energy tensor of the matter and ghost action $S_{m+g}$. Hence condition (2) is also satisfied and from $T_{--} = 0$ and the identity $\p_-^3 \mu \equiv  \( \p_+ - \mu \p_- \) \{f,x^-\} - 2  \{f,x^-\} \p_- \mu$ we recover the equation of motion of $\mu$
  \begin{align}
  \p_-^3 \mu - n^2 \p_- \mu = 0.
  \end{align}

We now turn to the condition (3) which requires $L = \Vevm{t_{++} - \mu^2 t_{--}}$. We would like to show that $t_{++} - \mu^2 t_{--}$ is the only combination of weight (2,0) operators that leads to a Virasoro algebra. Let us consider the generating functional of connected diagrams of the matter plus ghost action $S_{m+g}$. In two dimensions this is given by the Polyakov action~\cite{Polyakov:1987zb} which may also be written as
  \begin{align}
  & S_p = -\frac{c_M}{96 \pi} \iint |\g|^{\frac{1}{2}}|\g'|^{\frac{1}{2}} \diff^2 x \diff^2 x' R(x) G(x,x') R(x'), \label{eq:polyakov} 
  \end{align}
where $\g_{\mu\nu}$ is the metric at the boundary and $G(x,x')$ is the scalar Green function satisfying
  \begin{align}
  |\g|^{\frac{1}{2}} \Box G(x,x') = \d^{(2)}(x-x'). \label{eq:greens}
  \end{align}
The solution to this equation is given by   
  \begin{align}
  G(x,x') = \frac{1}{4\pi}\log{|x^+ - x'^+||f(x) - f(x')|}, \label{eq:greens2}
  \end{align}
where we recall that $\mu = \p_+ f/ \p_- f$. 

We are interested in the two-point functions of the stress energy tensor, i.e. 
  \begin{align}
  \Vev{t_{\mu\nu}(x) t_{\rho\s}(x')} = \frac{(4\pi)^2}{\sqrt{|\g|}\sqrt{|\g'|}} \frac{\d^{2} S_{p}}{\d \g^{\mu\nu}(x) \d \g^{\rho\s}(x')}.
  \end{align}
It is convenient to write the Polyakov action as
  \begin{align}
  S_p = -\frac{c_M}{96\pi} \int |\g|^{\frac{1}{2}} \diff^2 x R(x) \vp(x), && \Box \vp = R, \label{eq:polyakov2}
  \end{align}
whereupon the second-order variation of the action, omitting variations that lead to contact terms in the two-point function, reads
  \begin{align}
  \d^2 S_p = -\frac{c_M}{48\pi} \int \diff^2 x \lb \d(|\g|^{\frac{1}{2}}R) \d \vp + \frac{1}{2}|\g|^{\frac{1}{2}}R \d^{2}\vp \rb,
  \end{align}
and $\d\vp$, $\d^2 \vp$ are obtained from the variation of the second equation in~\eqref{eq:polyakov2}. 

Modulo contact terms the two-point functions of the stress-energy tensor are then given by
  \begin{align}
  \Vev{t_{++} t_{++}'} &= \frac{c/2}{(x^+-x'^+)^4} + \frac{c/2(\p_+ f)^2(\p_+'f')^2}{(f-f')^4}, \\
  \Vev{t_{++} t_{--}'} &= \frac{c/2(\p_+ f)^2(\p_-'f')^2}{(f-f')^4}, \\
  \Vev{t_{--} t_{--}'} &= \frac{c/2(\p_- f)^2(\p_-'f')^2}{(f-f')^4},
  \end{align}
where primed quantities correspond to primed coordinates. We can thus identify the generator of Virasoro transformations of the matter and ghost action with
  \begin{align}
  L = \Vevm{t_{++} - \mu^2 t_{--}}, \label{eq:tmunu2}
  \end{align}
since its two-point function reproduces the appropriate anomalous term that leads to the Virasoro algebra
  \begin{align}
  \Vev{L(x^+)L(0)} = \frac{c_M/2}{(x^+)^4}. \label{eq:boundaryvirope}
  \end{align}
Thus the Brown-York stress-energy tensor matches the expectation value of the stress-energy tensor of the dual theory at the boundary. This is not too surprising since the Polyakov action in the dual theory guarantees that the stress-energy tensor is covariantly conserved and that its trace reproduces the quantum anomaly. In particular, having identified $L$ with the generator of Virasoro transformations of the matter and ghost fields, their central charge can be obtained in the bulk from the anomalous transformation of $L$ and is given by $c_M = 3l/2G$ in the semiclassical approximation.

The Polyakov action is invariant under Kac-Moody transformations where
  \begin{align}
  \d x^- = \ll(x), && \d_{\ll} \mu = (\p_+ - \mu\p_-) \ll + \ll \p_- \mu, \label{eq:deltamuboundary}
  \end{align}
and $\ll$ satisfies $\p_-^3 \ll -  n^2 \p_- \ll = 0$. Therefore the parameter $\ll$ may be written as in eq.~\eqref{eq:sl2rparameter}. The Noether currents, computed from eq.~\eqref{eq:noncovariantpolyakov} and normalized by a factor of $2\pi$, are then given by
  \begin{align}
  \tilde J_0^- &= \frac{k}{2n}(\p_-^2\mu - n^2 \mu) - \frac{1}{n} \mu \Pi, && & \tilde J_0^+ &= \frac{1}{n} \Pi\label{eq:lightconecftcurrents} \\ 
  \tilde J_{\pm}^- &= \frac{k}{2n} e^{\pm nx^-}\( \p_-^2\mu \mp n\p_-\mu  \) - \frac{1}{n} \mu \Pi, && & \tilde J_{\pm}^+ &= \frac{1}{n}e^{\pm nx^-} \Pi \label{eq:lightconecftcurrents2}
  \end{align}
where the upper indices are spacetime indices and we have defined $\Pi = -\frac{k}{2}\(\{f,x^-\} + \frac{n^2}{2}\)$ to simplify these expressions. The Noether currents of the matter plus ghost action $j_a^i$ can be determined as follows. In the absence of the Polyakov action these symmetries are gauged at the classical level since they correspond to a combination of diffeomorphisms and Weyl transformations introduced by making the metric dynamical. Therefore the $\SL2R$ currents must vanish on-shell. Since the lightcone parameter $\mu$ enters as a Lagrange multiplier that sets $t_{--} = 0$ (see eq.~\eqref{eq:perturbation}), the currents must be proportional to $t_{--}$. Furthermore, since the Kac-Moody algebra is left-moving, we expect $j_a^-$ to have weight 1 under Virasoro transformations and, from the conservation equation, that $j_a^+$ has weight 0. Since $t_{--}$ has weight 0 as well, we have
  \begin{align}
  j_a^- &\propto \frac{1}{n} \mu t_{--}, && & j_a^+ \propto \frac{1}{n} t_{--}.
  \end{align}
The relative sign between these currents and those given in eqs.~\eqref{eq:lightconecftcurrents},~\eqref{eq:lightconecftcurrents2} is determined from the variation of the full action,
  \begin{align}
  \d S = \d S_{m+g} + \frac{k}{16\pi} \d S_p = \int \diff^2 x \lB \d\phi_a [\phi_a] + \d \mu \lb t_{--} - \frac{k}{2} \( \{f,x^-\} + \frac{n^2}{2} \)\rb \rB,
  \end{align}
where $[\phi_a]$ represents the equations of motion of the matter and ghost fields $\phi_a$. From the non-linear part of $\d\mu$ in eq.~\eqref{eq:deltamuboundary} we have 
  \begin{align}
  \d S = \int  \diff^2 x \, \p_- \lb \ll \mu\( t_{--} + \Pi \) \rb - \p_+ \lb\ll \( t_{--} + \Pi \) \rb + \dots
  \end{align}
where we see that both $t_{--}$ and $\Pi$ contribute with the same coefficient to the Kac-Moody currents. Thus,
  \begin{align}
  j_a^- = -\frac{1}{n} \mu t_{--}, && & j_a^+ = \frac{1}{n} t_{--}, \label{eq:mattersl2rcurrents}
  \end{align}
and, using the equations of motion, the Kac-Moody currents of the matter, ghost, and Polyakov actions, $J_a = j_a + \tilde J_a$ are given by
  \begin{align}
  & J_0^- = \frac{k}{2n} \( \p_-^2 \mu - n^2 \mu \), && & J_0^+ = 0, \\
  & J_{\pm}^- = \frac{k}{2n} e^{\pm nx^-}\( \p_-^2\mu \mp n\p_-\mu  \), && & J_{\pm}^+  = 0.
  \end{align}
Hence, the $J_a^-$ components match the Kac-Moody currents found in the bulk~\eqref{eq:sl2rcurrents}. That these currents generate an $\SL2R$ current algebra with level $k = -c_M/6$~\eqref{eq:sl2rOPE} was originally found in refs.~\cite{Polyakov:1988ok,Alekseev:1988ce,Alekseev:1990mp} and follows from the parametrization of $\mu$ given in eq.~\eqref{eq:musolution} and the transformation of $\mu$ given in eq.~\eqref{eq:deltamuboundary}.

The dual theory is also invariant under Virasoro transformations where,
  \begin{align}
  \d x^+ = \e(x^+), && \d x^- = -\frac{\e'(x^+)}{n} , && \d \mu = \p_+ \(\e\mu\) -\frac{\e''(x^+)}{n} - \frac{\e'(x^+)}{n} \p_- \mu ,
  \end{align}
Following arguments similar to those used in the derivation of the Kac-Moody currents, we find that on-shell, the Virasoro current $H^{\mu}$, where $\mu$ is a spacetime index, is given by
  \begin{align}
  H^- = t_{++} - \mu^2 t_{--} + T_s, &&  H^+ = 0, \label{eq:lightconevir}
  \end{align}
where $T_s$ is the twisted Sugawara tensor of eqs.~\eqref{eq:twisted1},~\eqref{eq:twisted2} with central charge $c_p =  6k$. Hence in the semiclassical limit the total central charge $c_{total} = c_M + c_p$ vanishes, in agreement with the computation in the bulk. In the quantum theory we expect $T_s$ to be given by,
  \begin{align}
  T_s = \frac{1}{k+2} \eta^{ab} J_a J_b + \p_+ J_0,
  \end{align}
where $J_a$ are the Kac-Moody currents of eq.~\eqref{eq:sl2rcurrents} and the central charge becomes
  \begin{align}
  c_p = \frac{3k}{k+2} + 6k.
  \end{align}
Then the OPEs between the Sugawara tensor and the Kac-Moody currents follow directly from the $\SL2R$ algebra and agree with the expressions found in the bulk.  

As discussed in Section~\ref{se:lightconegauge} the appearance of the twisted Sugawara tensor in the generator of Virasoro transformations agrees with the gravitational constraints on the dual theory. Indeed, from eq.~\eqref{eq:lightconetmunu} we have\footnote{Note that these equations hold for any value of $n$.}
  \begin{align}
  0 = T_{+-} \propto J_-, && 0 = T_{++} = t_{++} - \mu^2 t_{--} + T_s + \O(J_-).  \label{eq:constraints}
  \end{align}
which constraints $J_-$ to be an operator of weight 0 and that all physical states are singlets under Virasoro transformations. The first constraint is consistent with the appearance of the the twisted Sugawara tensor in the generator of Virasoro transformations, while the second constraint requires that physical states are singlets of the Virasoro symmetry. This result can be recovered in the bulk and is a consequence of the vanishing $\xi^r$ component of the asymptotic Killing vector~\eqref{eq:killing_vir}. A massive scalar field in AdS$_3$ with boundary conditions given by eq.~\eqref{eq:lightconebc} obeys the same asymptotic behavior as AdS$_3$ with standard Brown-Henneaux boundary conditions~\eqref{eq:bulkscalar}. Thus, under Virasoro transformations a scalar field in the bulk transforms asymptotically as
  \begin{align}
  \L_{\e} \phi \sim r^{\dd_-} \( \e\, \p_+\a - \frac{1}{n} \e'\p_- \a\) + r^{\dd_+} \( \e\, \p_+\b - \frac{1}{n} \e'\p_- \b\),
  \end{align}
where we can see that the Virasoro weight of all scalar operators vanishes. On the other hand, under Kac-Moody transformations generated by $J_0$, i.e. by the Killing vector~\eqref{eq:killing_sl2r} with $\ll = \l_0(x^+)$, we have
  \begin{align}
  \L_{\e} \phi \sim r^{\dd_-} \( \frac{\dd_-}{2} \l_0 \a + \l_0 \p_- \a\) + r^{\dd_+} \(\frac{\dd_+}{2} \l_0 \b + \l_0 \p_- \b\), \label{eq:sl2rweight}
  \end{align}
from which we conclude that the $\SL2R$ weight is given by $\l = h = \dd_+/2$ in agreement with the semiclassical limit of the result obtained in ref.~\cite{Polyakov:1987zb}\footnote{Note that in the case $n = 0$ we have $\ll = \l_0(x^+) x^-$ but also $\z^r = -r \p_- \ll /2$ for the r-component of the Kac-Moody Killing vector. Hence eq.~\eqref{eq:sl2rweight} is independent of $n$ and we recover the result $\l = h$ of~\cite{Polyakov:1987zb} when $n = 0$.},
  \begin{align}
  \l = h + \frac{\l(1-\l)}{k+2}.
  \end{align}

As in the conformal gauge studied in Section~\ref{se:conformaldualtheory}, we expect the theory to be non-unitary, in this case because $\SL2R$ does not have unitary representations (see e.g.~\cite{Dixon:1989cg}). Note however that the $\SL2R$ algebra is broken since the stress-energy tensor is twisted and the $\SL2R$ currents are not operators of weight one~\eqref{eq:virsl2rOPE3}; furthermore, note that the symmetry is realized non-linearly~\eqref{eq:virsl2rOPE3}. Hence the standard arguments from representation theory used to establish the non-unitarity of the theory do not apply. Nevertheless, even if the theory is not unitary we have constraints that can in principle be used to remove states with negative norm from the spectrum. We leave a detailed analysis of the unitarity of the theory 
for future work.
  
It is important to note that unlike Section~\ref{se:conformaldualtheory} where we considered the conformal gauge, in the lightcone gauge the gravitational degree of freedom $\mu$ couples to matter and ghosts~\eqref{eq:perturbation}. Hence it is instructive to consider a couple of examples that realize some of the results obtained in this section. The first example is a free scalar field whose action is given by
  \begin{align}
  S_m = \frac{1}{2}\int\sqrt{|\g|} \diff^2 x \g^{\mu\nu}\p_{\mu}\phi\p_{\nu}\phi = \int \diff^2 x\( -\p_+ \phi \p_- \phi + \mu \p_-\phi\p_-\phi\).  \label{eq:freescalar}
  \end{align}
The equation of motion reads
  \begin{align}
  0 = \p_-  \nabla_+ \phi, && \nabla_+ = \p_+ - \mu\p_-. \label{eq:eom1}
   \end{align}
and it is not difficult to show that $t_{--} = 2\pi(\p_- \phi \p_- \phi )$ obeys eq.~\eqref{eq:eomconstraint} since it is a consequence of covariant conservation of the stress-energy tensor. The action is left invariant by the $\SL2R$ transformations
  \begin{align}
  & \d x^- = \ll(x), && \d \phi = \ll \p_-\phi, && \d \mu = (\p_+ - \mu \p_-) \ll + \ll \p_- \mu, \label{eq:sl2r_mu}
  \end{align}
where $\ll(x)$ is given by eq.~\eqref{eq:sl2rparameter}. The Noether currents $j_a^{\mu}$, where $\mu$ is a spacetime index, are given by
  \begin{align}
  j_a^{-} = - \frac{2\pi}{n} \mu\p_-\phi\p_-\phi, &&   j_a^+ = \frac{2\pi}{n} \p_-\phi\p_-\phi.
  \end{align}
and agree with the expression obtained under general arguments~\eqref{eq:mattersl2rcurrents}.

The action~\eqref{eq:freescalar} is also invariant under the $\d x^+ = \e(x^+)$ reparametrizations where the fields transform as
  \begin{align}
  & \d \phi = \e \p_+ \phi, & \d \mu = \p_+\e\mu + \e\p_+\mu. \label{eq:vir_mu}
  \end{align}
Hence the action is invariant under Virasoro transformations where $\d x^+ = \e(x^+)$, $\d x^- = -1/n \e'(x^+)$ and $\phi$ transforms in the obvious way. When the Polyakov action is included, the full generator of Virasoro transformations, $H^{\mu}$, is given by
  \begin{align}
  H^- = t_{++} - \mu^2 t_{--} + T_s, && H^+ = 0.
  \end{align}
Here we identify $L = t_{++} - \mu^2 t_{--}$ with the generator of Virasoro transformations of the free scalar field. Using the expressions for the stress-energy tensor $t_{\mu\nu}$ of the free scalar theory, $L$ can be more conveniently written as,
  \begin{align}
  L = t_{++} - \mu^2 t_{--} = 2\pi (\nabla_+ \phi)^2,
  \end{align}
which is a left-moving operator thanks to the equations of motion of $\phi$. We can also check that its OPE contains the appropriate $1/(x^+)^4$ term found in eq.~\eqref{eq:boundaryvirope}. To see this note that the $\phi(x^+)\phi(0)$ OPE can be obtained from the Green function given in eq.~\eqref{eq:greens2}. Therefore the OPE of $\nabla_+\phi$ is given by
  \begin{align}
  \nabla_+ \phi(x) \nabla_+ \phi(0) = -\frac{1}{4\pi}\frac{1}{x^+},
  \end{align}
and it is readily seen that $L$ obeys the expected OPE
  \begin{align}
  L(x^+) L(0) = \frac{1/2}{(x^+)^4} + \frac{2 L(0)}{(x^+)^2} + \frac{\p_+ L(0)}{x^+}.
  \end{align}

As a second example let us consider a free Majorana fermion in the lightcone gauge,
  \begin{align}
  S = \frac{1}{2} \int \sqrt{|\g|}\diff^2x\, i \bar \psi \g^{\mu} e_{a}^{\phantom{a}\mu} \p_{\mu} \psi = \int \diff^2 x \,\psi_- \( \p_+ - \mu \p_- \) \psi_- + \psi_+ \p_- \psi_+,
  \end{align}
where $\bar \psi = i \psi^{\dagger} \g^0$, $\psi^{\dagger} = \( \psi_-\,\psi_+\)$, and the gamma matrices and zweibein are given by
  \begin{align}
  \g^0 = \(\begin{array}{rrr}
  			0 && -1 \\
			1 && 0 \end{array} \), && \g^1 = \(\begin{array}{rrr}
			 							0 && 1 \\
										1 && 0 \end{array} \), &&  e^{a}_{\phantom{a}\mu} = \( \begin{array}{rrr}
										1 && \mu \\
										0 && 1 \end{array} \).
  \end{align}
The equations of motion are readily seen to be
  \begin{align}
  \p_- \psi_+ = 0, && (\p_+ - \mu \p_-) \psi_- - \frac{1}{2} \p_- \mu\,\psi_- = 0,
  \end{align}
and it is easy to check that $t_{--} = -\pi(\psi_-\p_-\psi_-)$ satisfies eq.~\eqref{eq:eomconstraint}.

The action is invariant under $\SL2R$ transformations $\d x^- = \ll(x)$ where
  \begin{align}
  \d \psi_+ = 0, && \d \psi_- = \ll \p_- \psi + \frac{1}{2} \p_- \ll \psi, && \d \mu = (\p_+ - \mu \p_-) \ll + \ll \p_- \mu.
  \end{align}
Using the equations of motion, the corresponding Noether currents are given by
  \begin{align}
  j_a^{-} = - \frac{\pi}{n} \mu\psi_-\p_-\psi_- &&   j_a^+ = \frac{\pi}{n} \psi_-\p_-\psi_-,
  \end{align}
in agreement with eq.~\eqref{eq:mattersl2rcurrents}. The action is also invariant under the chiral transformations $\d x^{+} = \e(x^+)$ where the fields transform as
  \begin{align}
  \d \psi_+ = \e \p_+ \psi_+ + \frac{1}{2} \p_+ \e \psi_+, && \d \psi_- = \e \p_+ \psi_-, && \d \mu = \p_+\e \mu + \e\p_+\mu.
  \end{align}
Hence the Virasoro transformation parametrized by $\d x^+ = \e(x^+)$ and $\d x^- = -1/n \e'(x^+)$ is also a symmetry of the theory. As for the free scalar field the full generator of Virasoro transformations, which includes contributions from the Polyakov action, is given by
  \begin{align}
  H^- = t_{++} - \mu^2 t_{--} + T_s, && H^+ = 0,
  \end{align}
where we identify $L = t_{++} - \mu^2 t_{--}$ with the generator of Virasoro transformations of the matter and ghost fields. For the free fermion theory $L$ is given by
  \begin{align}
  L = t_{++} - \mu^2 t_{--} = -\pi \psi_+ \p_+ \psi_+,
  \end{align}
which is the $T_{++}$ component of the stress-energy tensor of a free fermion field in a flat background. Since the equation of motion of $\psi_+$ does not change in the lightcone gauge, it is readily seen that $L$ is conserved and that the OPE $L(x^+)L(0)$ is given by
  \begin{align}
  L(x^+) L(0) = \frac{1/4}{(x^+)^4} + \frac{2 L(0)}{(x^+)^2} + \frac{\p_+ L(0)}{x^+}.
  \end{align}
  % 

%%%%%%%%%%%%%%%%%%%%%%%%%%%%%%%%%%%%%%%%%%%%%%%%%%%
\section{Conclusions}
%%%%%%%%%%%%%%%%%%%%%%%%%%%%%%%%%%%%%%%%%%%%%%%%%%%
\label{se:conclusions}

In this paper we have studied the recently-proposed free boundary conditions for AdS$_3$ and shown they reproduce many results of 2D quantum gravity in the conformal and lightcone gauges. An interesting aspect of these boundary conditions is that they enhance the asymptotic symmetry group of AdS$_3$ and support both Virasoro and Kac-Moody algebras. In particular, we have shown that the Virasoro and Kac-Moody charges of the bulk theory match the corresponding quantities at the boundary. We have seen that the appropriate generator of Virasoro transformations has a vanishing central charge, in agreement with the dual theory. This generator is also consistent with the constraints imposed on the dual theory, which require that physical states are singlets under Virasoro transformations.

It is interesting to note that the Kac-Moody currents depend only on the boundary data, i.e. on the gravitational degrees of freedom, $\vp$ or $\mu$. Hence BTZ solutions have vanishing Kac-Moody charges and the standard form of Cardy's formula can be used to match the entropy of BTZ black holes. In particular, the lightcone gauge admits only extremal BTZ black holes provided $n = 0$. This is consistent with the single copy of the Virasoro algebra found at the boundary. 

This paper is a first step towards more rigorous checks of the correspondence and it would be interesting to explore other aspects of 2D quantum gravity from holography. Here we only considered a boundary with either a cylinder or a torus topology. Since the new asymptotic symmetries defined by the new boundary conditions studied here act only on the gravitational sector and two-dimensional gravity is topological, an obvious interesting avenue for future research is to extend our study to boundaries with more complex topology.  Another interesting question is to understand if any relation exists between boundary conditions with dynamical gravity and a near-horizon description of four-dimensional Schwarzschild or Kerr black holes.

%%%%%%%%%%%%%%%%%%%%%%%%%%%%%%%%%%%%%%%%%%%%%%%%%%%
\section*{Acknowledgments}
%%%%%%%%%%%%%%%%%%%%%%%%%%%%%%%%%%%%%%%%%%%%%%%%%%%

It is a pleasure to thank Sergei Dubovsky and Jihun Kim for useful discussions, C\'edric Troessaert for clarifying certain aspects of his work, and Hovhannes Grigoryan for comments on the manuscript. We would also like to thank Matthew Headrick for making his Mathematica package for tensor algebra freely available online\footnote{\url{http://people.brandeis.edu/~headrick/Mathematica/}}, and Geoffrey Comp\`ere for useful comments. M.P. is supported in part by NSF grants PHY-0758032, PHY-1316452. M.P thanks the ERC Advanced Investigator Grant No.\@ 226455 {\em Supersymmetry, Quantum Gravity and Gauge Fields (Superfields)} for support and CERN for hospitality during the completion of this paper.

%%%%%%%%%%%%%%%%%%%%%%%%%%%%%%%%%%%%%%%%%%%%%%%%%%%

\end{document}